\documentclass[
    11pt,
    reprint,
 	secnumarabic,
 	superscriptaddress,
 	numerical,
 	amssymb,
 	notitlepage,
	frontmatterverbose, 
 	amsmath,
 	aps,
	prstper,
	floatfix,
    longbibliography
]{revtex4-1}

\usepackage{mathrsfs}

\usepackage{ulem} 
\usepackage{graphicx} 
\usepackage[]{natbib} 
\usepackage[usenames,dvipsnames]{xcolor} 
\usepackage{dcolumn} 
\usepackage{bm} 
\usepackage{placeins} 
\usepackage[mathscr,mathcal]{euscript} 
\usepackage{epstopdf} 
\usepackage{nameref} 

\usepackage[urlcolor=blue, breaklinks=true, hyperindex, colorlinks, bookmarks=true,linkcolor=black,hidelinks,citecolor=blue]{hyperref}

\usepackage{makecell}
\usepackage{boldline}

\setcellgapes{3pt}
\usepackage{adjustbox}
\usepackage{multirow}

\newcolumntype{?}{!{\vrule width 1pt}}

\newcommand{\boldit}[1]{\boldsymbol{#1}}
\renewcommand{\vec}[1]{\boldsymbol{#1}}

\begin{document}


\title[]{
How optical excitation controls the structure and properties of vanadium dioxide
}

\author{Martin R. Otto}
 \email{martin.otto@mail.mcgill.ca} 
 \affiliation{
Department of Physics, Center for the Physics of Materials, McGill University, 3600 University Street, Montreal, QC, CA 
}%

\author{Laurent P. Ren\'{e} de Cotret} 
 \affiliation{
Department of Physics, Center for the Physics of Materials, McGill University, 3600 University Street, Montreal, QC, CA 
}%

\author{David A. Valverde-Chavez} 
 \affiliation{
Department of Physics, Center for the Physics of Materials, McGill University, 3600 University Street, Montreal, QC, CA 
}%

\author{Kunal L. Tiwari} 
 \affiliation{
Department of Physics, Center for the Physics of Materials, McGill University, 3600 University Street, Montreal, QC, CA 
}%

\author{Nicolas \'Emond} 
 \affiliation{
Institut National de la Recherche Scientifique, Centre \'Energie Mat\'eriaux et T\'el\'ecommunications, Universit\'e du Qu\'ebec, Varennes, QC J3X 1S2, Canada 
}%

\author{Mohamed Chaker} 
 \affiliation{
Institut National de la Recherche Scientifique, Centre \'Energie Mat\'eriaux et T\'el\'ecommunications, Universit\'e du Qu\'ebec, Varennes, QC J3X 1S2, Canada 
}%

\author{David G. Cooke} 
 \affiliation{
Department of Physics, Center for the Physics of Materials, McGill University, 3600 University Street, Montreal, QC, CA 
}%

\author{Bradley J. Siwick}
 \email{bradley.siwick@mcgill.ca}
 \affiliation{
Department of Physics, Center for the Physics of Materials, McGill University, 3600 University Street, Montreal, QC, CA 
}%
 \affiliation{
Department of Chemistry, McGill University, 801 Sherbrooke Street W, Montreal, QC, CA 
}%

\date{\today}



\begin{abstract}
We combine ultrafast electron diffraction and time-resolved terahertz spectroscopy measurements to unravel the connection between structure and electronic transport properties during the photoinduced insulator-metal transitions in vanadium dioxide. We determine the structure of the metastable monoclinic metal phase, which exhibits anti-ferroelectric charge order arising from a thermally activated, orbital-selective phase transition in the electron system. The relative contribution of this photoinduced monoclinic metal (which has no equilibrium analog) and the photoinduced rutile metal (known from the equilibrium phase diagram) to the time and pump-fluence dependent multi-phase character of the film is established, as is the respective impact of these two distinct phase transitions on the observed changes in terahertz conductivity. Our results represent an important new example of how light can control the properties of strongly correlated materials and elucidate that multi-modal experiements are essential when seeking a detailed connection between ultrafast changes in optical-electronic properties and lattice structure in complex materials.
\end{abstract}

\maketitle

The insulator-metal transition (IMT) in vanadium dioxide (VO$_2$) is a benchmark problem in condensed matter physics~\cite{Morin1959,GOODENOUGH1971,Zylbersztejn1975,Eyert2002,BasovRMP2011,budai2014}, as it provides a rich playground on which lattice-structural distortions and strong electron correlations conspire to determine emergent material properties. The equilibrium phase diagram of pure VO$_2$ involves a high-temperature tetragonal (rutile, $R$) metal that is separated from several structurally distinct low-temperature insulating phases (monoclinic $M_1$, $M_2$ and triclinic $T$) depending on pressure or lattice strain. The transition to these lower-symmetry insulating phases occurs in the vicinity of room temperature and is sensitive to doping (Cr and W), making VO$_2$ interesting for a range of technological applications~\cite{Hendaoui2013,Ryckman2012,Miller2017}. Since its discovery there has been a lively discussion in the literature about the driving force responsible for the IMT in VO$_2$ and the nature of the insulating and metallic phases that has revolved around the role and relative importance of electron-lattice and electron-electron interactions. The stark dichotomy between Peierls~\cite{GOODENOUGH1971} and Mott~\cite{Mott1974} pictures characterizing the earliest explanations have recently given way to a nuanced view that the insulating phases of VO$_2$ are non-standard Mott-Hubbard systems where both electron-lattice and electron-electron interactions play important roles in determining the electronic properties of all the equilibrium phases~\cite{Eyert2011,Weber2012,Biermann2005,Brito2016,Huffman2017,Najera2018}.   

Photoexcitation using ultrafast laser pulses has provided another route to initiate the transition between the insulating and metallic phases of VO$_2$ since it was discovered that the IMT occurs very rapidly following femtosecond laser excitation with sufficient fluence~\cite{Becker1994}. Since this discovery, VO$_2$ has been the focus of many time-resolved experiments including X-ray~\cite{Cavalleri2001,Hada2011} and electron~\cite{Baum2007,Morrison2014a,tao2016} diffraction, X-ray absorption~\cite{Cavalleri2005,Haverkort2005}, photoemission~\cite{Wegkamp2014} and optical spectroscopies~\cite{Cavalleri2004,Kubler2007,Pashkin2011,wall2012,Cocker2012,Wall2013,Mayer2015,Jager2017,Abreu2017} from terahertz to ultraviolet aimed at uncovering the connection between the photo-induced IMT and changes in lattice structure.

\begin{figure*}
	\includegraphics[width = 0.85\textwidth]{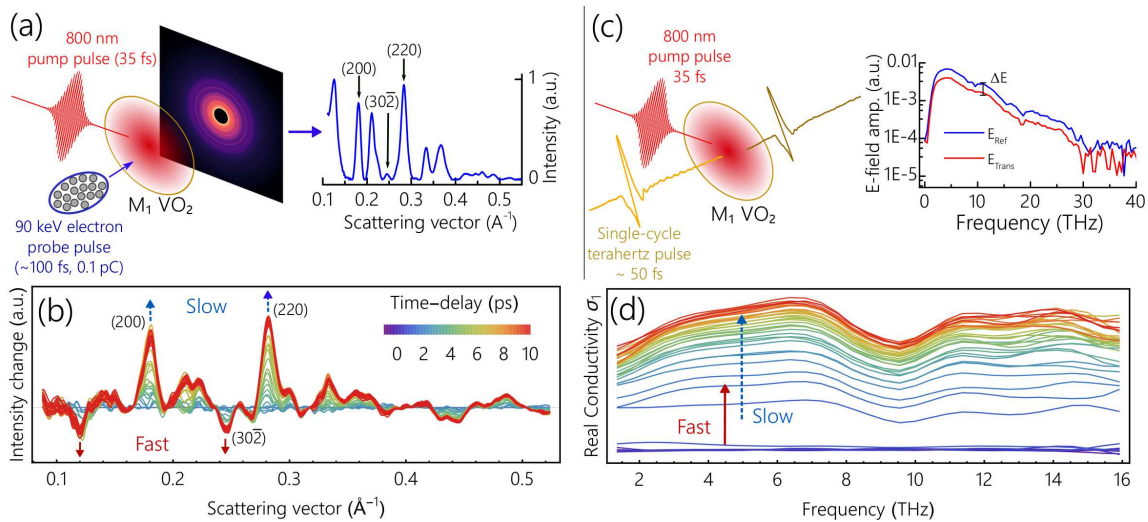}
	\caption{Multi-modal pump-probe experiments performed on the photoinduced phase transtions in VO$_2$ (a) Ultrafast electron diffraction.  The 50~nm thick polycrystalline VO$_2$ sample is photoexcited by a 35~fs 800~nm optical pump pulse. The probe electron pulse has a temporal full-width at half-maximum of $\sim$100~fs and arrives after a precisely controlled time delay. Debye-Scherrer patterns are collected by a charge-coupled device (CCD) camera for each time-delay value. A typical azimuthally-averaged diffraction pattern is shown to the right, identifying the Bragg peaks for VO$_2$ most discussed in this work; (200), (220) and (30$\bar{2}$). (b) The pump-induced difference in diffracted intensity up to 10~ps that follows photoexcitation at 23~mJ/cm$^2$. Blue and red arrows indicate important aspects of the slow and fast processes respectively. (c) Time-resolved terahertz (THz) spectroscopy. The electric field waveform of the transmitted, single cycle multi-THz pulse with spectral components spanning 1--30 THz is measured in the absence, $E_{\textup{ref}}(t)$, and presence, $E_{\textup{pump}}(t,\tau)$ of a photoexcitation at pump-probe delay time $\tau$. The differential $\Delta E(t,\tau)=E_{\textup{ref}}(t)-E_{\textup{pump}}(t,\tau)$ and $E_{\textup{ref}}(t)$ are Fourier transformed and the resulting amplitude and phase spectra are used to extract the dynamically changing complex optical conductivity $\tilde{\sigma}(\omega,\tau)=\sigma_1(\omega,\tau)+i\sigma_{2}(\omega,\tau)$ (d) Change in the real part of the optical conductivity ($\sigma_1(\omega,\tau)$) from 1--16 THz computed from the differential THz waveforms.} 
	\label{FIG:pump-probe}
\end{figure*}

Recently, ultrafast electron diffraction (UED) and mid-infrared spectroscopy were combined to show that there are two distinct photo-induced IMTs in $M_1$ VO$_2$~\citep{Morrison2014a}. The first, accessible at relatively high pump fluence, is an analog of the equilibrium IMT and is associated with the lattice-structural transition between $M_1$ and $R$ crystallography expected from the equilibrium phase diagram. The second, accessible at lower pump fluence, has no equilibrium analog and yields a metastable, structurally distinct monoclinic metal phase ($\mathscr{M}$) that retains the crystallographic symmetry of its parent equilibrium monoclinic phase.  Here, UED and time-resolved terahertz spectroscopy (TRTS) are combined (Fig.~\ref{FIG:pump-probe}) to perform a detailed structure-property correlative study of the photo-induced phase transitions in VO$_2$.  UED measurements are used to determine the real-space structure of the transient metastable $\mathscr{M}$ phase (Sec.~\ref{sec:structure}). The fluence dependence of the $\mathscr{M}$ and $R$ phase fractions that form the heterogeneous multi-phase specimen following photoexcitation are measured and shown to directly correspond with the TRTS measurements (Sec.~\ref{sec:fluence_dependence}). From this correspondence we determine the low-frequency terahertz conductivities of both photo-induced $\mathscr{M}$ and $R$ phases. The kinetics of the $M_1\rightarrow\mathscr{M}$ phase transition is consistent with thermal activation driven by electron temperature with an activation energy of $304\pm109~$meV (Sec.~\ref{sec:kinetics}). Information on the free energy landscape in VO$_2$ and its dependence on structural distortions and orbital occupancy is also discussed. Our results provide a unified view of the photo-induced structural phase transitions in VO$_2$ and their relationship to changes in the low-frequency terahertz conductivity.   


\section{Experimental results}\label{sec:results}
\begin{figure}
	\centering
    \includegraphics[width = 0.4\textwidth]{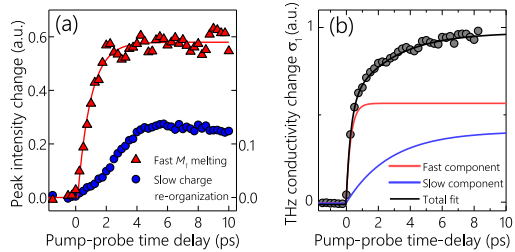}
    \caption{Comparison of ultrafast electron diffraction and time-resolved terahertz spectroscopy measurements. (a) The red triangles show the suppression of the (30$\bar{2}$) peak ($\sim350~$fs, red line)  associated with the $M_1$--$R$ transition (see Figure~\ref{FIG:pump-probe}). The blue circles show the intensity increase of the (200) and (220) peaks, which are associated with the slow re-organization ($\sim 2~$ps) and the formation of the $\mathscr{M}$ phase. (b) The transient change in THz conductivity (spectrally integrated from 2--6~THz, shown as grey circles) is well-described by a bi-exponential function comprised of both fast ($\sim350~$fs, red line) and slow ($\sim2.5~$ps, blue line) time constants which are in quantitative agreement with those of the structural transitions observed with UED.}
    \label{FIG:ued_thz_time_traces}
\end{figure}

UED measurements of pulsed laser deposited VO$_2$ films (50 nm) reveal rich pump-fluence dependent dynamics up to the damage threshold of $\sim$40~mJ/cm$^2$ (35 fs, 800 nm, $f_{\textup{rep}}=50-200~$Hz). Figure~\ref{FIG:pump-probe}~(a) shows a typical one-dimensional powder diffraction pattern for equilibrium VO$_2$ in the $M_1$ phase and identifies the (200), (220) and (30$\bar{2}$) peaks. The (30$\bar{2}$) peak acts as an order parameter for the $M\rightarrow R$ transition, since it is forbidden by the symmetry of the $R$ phase, while (200) and (220) peaks are present in all equilibrium phases. Consistent with previous work~\citep{Morrison2014a}, the pump-induced changes to diffracted intensity (Fig.~\ref{FIG:pump-probe}~(b), 23 mJ/cm$^2$) indicate two distinct and independent photo-induced structural transformations. The first is a rapid ($\tau_{(30\bar{2})}\approx 300~$fs) non-thermal melting of the periodic lattice distortion (dimerized V--V pairs) present in $M_1$, evident in Fig.~\ref{FIG:pump-probe}~(b) and Fig.~\ref{FIG:ued_thz_time_traces}~(a) as a suppression of the ($30\bar{2}$) and related peak intensities. The second is a slower ($\tau_{(200),(220)}\approx 2~$ps) transformation associated with a significant increase in the intensity of the (200), (220) and other low index peaks whose time-dependence is also shown in Fig.~\ref{FIG:pump-probe}~(b) and Fig.~\ref{FIG:ued_thz_time_traces}~(a). As we will show, at low pump fluences ($\sim$3--8 mJ/cm$^2$) the slow process is exclusively observed, while at high pump fluences ($>35~$mJ/cm$^2$) the fast process dominates; these are independent structural transitions, the slow process does not follow the fast process.  

Complementary TRTS measurements were performed on the same samples under identical excitation conditions to determine the associated changes in the time-dependent complex conductivity, $\tilde{\sigma}(\omega,\tau)$ (see Fig.~\ref{FIG:pump-probe}~(c) and (d)). The pump-induced changes in real conductivity, $\sigma_1(\omega,\tau)$, over the $\sim$2--20 THz frequency range (Fig.~\ref{FIG:pump-probe}~(d)) also exhibit fast ($\Delta\sigma_1^{\textup{fast}}$) and slow ($\Delta\sigma_1^{\textup{slow}}$) dynamics, consistent in terms of timscales and fluence dependence with those described above for the UED measurements and similar measurements performed on sputtered VO$_2$ films in the 0.5--2 THz window~\cite{Cocker2012}. Additional structure at higher frequencies is due to optically active phonons associated with O--cage vibrations around V atoms~\cite{Kubler2007}.  We connect the observed THz response to the two structural transformations by focusing on the integrated spectral region from 2--6~THz which includes exclusively electronic contributions to the conductivity (Drude-like) and omits phonon resonances~\cite{Kubler2007,Pashkin2011,wall2012}. Figure~\ref{FIG:ued_thz_time_traces}~(b) shows an example of the transient real conductivity measured at 22~mJ/cm$^2$ along with the fast and slow exponential components plotted individually. We find that these time constants are in excellent agreement with those of the fast and slow processes determined from the UED measurements (Fig.~\ref{FIG:ued_thz_time_traces}). This correspondence holds over the enitre range of fluences investigated. 

\subsection{Structure of the monoclinic metallic phase}\label{sec:structure}

Since its discovery~\citep{Morrison2014a}, the structure of the photoinduced $\mathscr{M}$ phase and its relationship to the parent $M_1$ phase has remained unclear. Here we use measured UED intensities to determine the changes in the electrostatic crystal potential, $\Phi(\vec{x})$, associated with the transformation between $M_1$ and $\mathscr{M}$ phases. The centro-symmetry of the monoclinic and rutile phases provides a solution to the phase problem~\cite{Elsaesser2010} and allows for the reconstruction of the full three-dimensional real-space electrostatic potential from each one-dimensional diffraction pattern obtained using UED.

\begin{figure*}
	\centering
    \includegraphics[width = 0.8\textwidth]{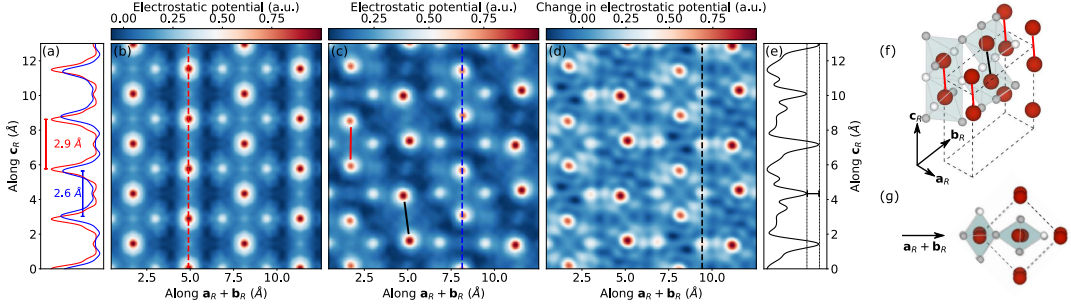}
    \caption{A real-space view of the photoinduced changes to $\Phi(\boldit{x})$ in VO$_2$. The plane of interest is spanned by $\boldit{c}_{R}$ and $\boldit{a}_{R}+\boldit{b}_{R}$. (a) Line-cuts of $\Phi(\boldit{x})$ along the red and blue dashed lines shown in (b) and (c) respectively. (b) $\Phi(\boldit{x})$ for $R$--phase VO$_2$ showing the absence of V--V dimerization and tilting. (c) $\Phi(\boldit{x})$ for $M_1$--phase VO$_2$ showing V--V dimerization and tilting illustrated by solid black (in-plane tilt) and red (out-of-plane tilt) lines which are also illustrated in the 3D structure ((f)). The line cuts in (a) show the expected undimerized V--V and dimerized V--V lengths of 2.9~\AA~and 2.6~\AA~respectively. (d) $\Delta\Phi(\boldit{x})$ for $\mathscr{M}$--phase VO$_2$ 10~ps after photoexcitation at a fluence of 6~mJ/cm$^2$. V--V dimerization from (c) is preserved and O atoms in the octahedra nearest to V atoms display an increase in $\Phi(\boldit{x})$ resulting in anti-ferroelectric order up the $\boldit{c}_{R}$ axis indicated by arrows. (e) Line-cut of $\Delta\Phi(\boldit{x})$ shown in (d) along the dashed black line which intersects a chain of O atoms. The anti-ferroelectric order is seen as an increase in $\Phi(\boldit{x})$ on alternating O atoms. (f) 3-dimensional structure of VO$_2$. V atoms appear as large red spheres and O atoms are grey (white) as per the anti-ferroelectric ordering depicted in (d) and (e). In-plane (out-of-plane) V--V dimers are connected by black (red) lines. (g) 3-dimensional structure of VO$_2$ looking down the $c_R$ axis.
    }
    \label{FIG:atomic_potential_maps}
\end{figure*}

Figure~\ref{FIG:atomic_potential_maps} shows slices of $\Phi(\vec{x})$ for VO$_2$ in the $R$ (b) and $M_1$ (c) phases obtained using this procedure. The slices shown are aligned vertically along the rutile $\vec{c}_R$ axis, and horizontally cut the unit cell along $\vec{a}_R + \vec{b}_R$ as indicated in the 3D structural model of $M_1$ VO$_2$ (Fig.~\ref{FIG:atomic_potential_maps}~(f)). In this plane, adjacent vanadium chains are rotated by 90$^{\circ}$, with dimers tilting either in or orthogonal to the plane of the page as indicated in Fig.~\ref{FIG:atomic_potential_maps}~(c) and (f). The lattice parameters obtained from these reconstructions are in excellent agreement with published values for the two equilibrium phases (Fig.~\ref{FIG:atomic_potential_maps}~(a)). The autocorrelation of $\Phi(\vec{x})$ is also in quantitative agreement with the Patterson function computed directly from the UED data (See Fig.~S6 in~\cite{SM}).

In Fig.~\ref{FIG:atomic_potential_maps}~(d) the changes in $\Phi(\vec{x})$ associated with the $M_1$--$\mathscr{M}$ transition are revealed. This map is computed from the measured $\Delta I_{\vec{G}}$ between the $\mathscr{M}$ and $M_1$ phases 10 ps after photoexcitation at 6 mJ/cm$^2$. The preservation of $M_1$ crystallography is clear; \textit{i.e.} V--V dimerization and tilting along the $\vec{c}_R$ axis. Also evident is the transition to a novel 1D anti-ferroelectric charge order along $\vec{c}_R$. In the equilibrium phases all oxygen atoms are equivalent, but in the $\mathscr{M}$ phase there is a periodic modulation in $\Phi(\vec{x})$ at the oxygen sites along the $\vec{c}_R$ axis indicated by arrows. This modulation is commensurate with the lattice constant (Fig.~\ref{FIG:atomic_potential_maps}~(c)-(e)). The oxygen atoms exhibiting the largest changes are those associated with the minimum V--O distance in the octahedra and, therefore, the V--V dimer tilt. This emphasizes the importance of the lattice distortion to the emergence of the $\mathscr{M}$ phase. The anti-ferroelectric lattice distortion in $M_1$ was already emphasized by Goodenough~\cite{GOODENOUGH1971} in his seminal work on VO$_2$. Significant changes in electrostatic potential are also visible between vanadium atoms in the octahedrally-coordinated chains along $\vec{c}_R$ that is consistent with a delocalization or transfer of charge from the V--V dimers to the region between dimers. All of these observations suggest that the $\mathscr{M}$ phase emerges from a collective reorganization in the electron system alone.    


\subsection{Fluence dependence}\label{sec:fluence_dependence}
\begin{figure*}
\centering
 	\includegraphics[width = 0.8\textwidth]{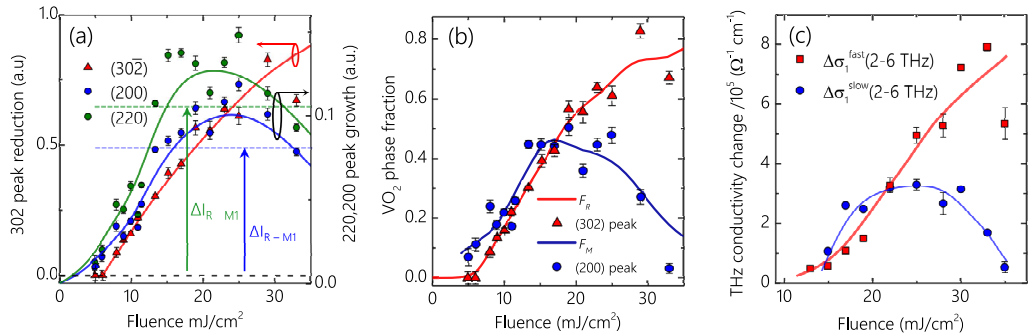}
    \caption{(a) Fluence dependence of diffraction peak intensity changes ($\Delta I_{hkl}$) determined by ultrafast electron diffraction. Fast non-thermal melting of $M_1\rightarrow~R$ crystallites occurs above a threshold fluence of $\sim 8~$mJ/cm$^2$ and yields a reduction of the (30$\bar{2}$) peak intensity (red triangles). Solid lines serve as a guide to the eye. The slower charge re-organization transition of $M_1\rightarrow~\mathscr{M}$ crystallites increases the (200) and (220) peak intensities (blue and green circles) which exhibit a maximum increase in the vicinity of 20~mJ/cm$^2$. At higher fluence the increase becomes less as the samples becomes predominantly $R$. The blue and green horizontal dashed lines represent the difference in the (200) and (220) peak amplitudes between the equilibrium $R$ and $M_1$ phases. (b) Phase volume fractions for the $\mathscr{M}$ (blue line) and $R$ (red line) phases calculated using the ultrafast electron diffraction data shown in (a) via the volume phase fraction model (see supplementary material~\cite{SM}). Red triangles are the ($30\bar{2}$) data points from (a) and blue circles are the (200) data points in (a) (scaled for clarity) with the $M_1\rightarrow R$ phase transition contribution subtracted. (c) Fluence dependence of the fast $\Delta\sigma_1^{\textup{fast}}$ and slow $\Delta\sigma_1^{\textup{slow}}$ components of the transient terahertz optical conductivity (red and blue respectively). Solid lines serve as a guide to the eye. $\Delta\sigma_1^{\textup{fast}}$ increased steadily and corresponds to the formation of $R$ crystallites and $\Delta\sigma_1^{\textup{slow}}$ attains a maximum at 25~mJ/cm$^2$ and corresponds to the formation of $\mathscr{M}$ crystallites. }
    \label{FIG:ued_thz_measurements}
\end{figure*}

We have established that there are two qualitatively distinct ultrafast photo-induced phase transtions in vanadium dioxide. The pump-fluence dependence of the sample response, specifically the heterogenous character of the film following photoexcitation (due to both $M_1\rightarrow\mathscr{M}$ and $M_1\rightarrow R$ transformations) and the corresponding changes in conductivity, is addressed in this section. UED intensities report on the fluence dependence of both structural phase transitions (Fig.~\ref{FIG:ued_thz_measurements}~(a)). As described earlier, the change in the ($30\bar{2}$) peak intensity provides an order parameter exclusively for the $M_1\rightarrow R$ transition, while the (200) and (220) peak intensities report on both $M_1\rightarrow\mathscr{M}$ and $M_1\rightarrow R$ transformations. Measurements of the (30$\bar{2}$) peak intensity (Fig.~\ref{FIG:ued_thz_measurements}~(a) red triangles) clearly demonstrate a fluence threshold of $\sim8~$mJ/cm$^2$ for the $M_1\rightarrow R$ transformation that is consistent with previous work~\cite{Morrison2014a,Baum2007,Cavalleri2001}. Above this threshold the suppression of the (30$\bar{2}$) peak increases approximately linearly with fluence up to a magnitude greater than $75\%$ at $\sim30$~mJ/cm$^2$, a result that is inconsistent with a ``two-step'' model that involves fast V--V dimer dilation followed by slow dimer rotation. Complete V--V dimer dilation yields a maximum (30$\bar{2}$) peak suppresion of $~50\%$. Instead, this data is consistent with a picture where the PLD of the $M_1$ phase simply melts in $\sim300~$fs.  The photoinduced fraction of $R$--phase VO$_2$ reaches $\sim75\%$ of the film on this timescale at the highest pump fluences reported. The (200) and (220) peaks (Fig.~\ref{FIG:ued_thz_measurements}~(a) blue and green circles) show a more complicated fluence dependence, reaching a maximum change in intensity in the 20~mJ/cm$^2$ range. At the highest excitation fluences reported the intensity changes in the (200) and (220) peaks correspond to the relative increase expected for the $R$ phase compared to the $M_1$ of VO$_2$. The maximum at $\sim20$~mJ/cm$^2$ is entirely due to the presence of the $\mathscr{M}$ phase as we demonstrate by converting the changes in UED intensities to phase volume fractions~\cite{SM}.

We denote $F_R$ as the phase volume fraction for the $R$ phase and $F_{\mathscr{M}}$ for the $\mathscr{M}$ phase. The results of the model are shown in Fig.~\ref{FIG:ued_thz_measurements}~(c). For fluences below the structural IMT fluence threshold of $\sim 8~$mJ/cm$^2$, we observe clearly that only a small percentage ($\sim 10\%$) of $\mathscr{M}$ crystallites have been formed by photoexcitation. As the fluence increases, the photoexcitation of $R$ crystallites begins at the threshold and increases roughly linearly afterwards. The $\mathscr{M}$ phase, achieves a maximum in the vicinity of 20~mJ/cm$^2$ (consistent with Fig.~\ref{FIG:ued_thz_measurements}~(a)) where we determine that $F_{\mathscr{M}}=45\pm13\%$. At greater fluences, $F_{\mathscr{M}}$ decreases as the material becomes increasingly $R$ phase due to stronger photoexcitation. The data points shown in Fig.~\ref{FIG:ued_thz_time_traces}~(c) as blue circles are an average of the (200) and (220) data points from (a) with the contribution from the $M_1\rightarrow R$ phase transistion subtracted.  

We obtain quantitatively consistent results for the fluence dependence of the transient conductivity obtained by TRTS, firmly establishing a link between the differential structure and differential electronic response. Figure~\ref{FIG:ued_thz_measurements}~(b) shows the fluence dependence of the fast ($\Delta\sigma_1^{\textup{fast}}$) and slow ($\Delta\sigma_1^{\textup{slow}}$) conductivity terms. The $\Delta\sigma_1^{\textup{fast}}$ component corresponds to the conductivity response associated with the transition from $M_1\rightarrow R$ as it increases steadily with fluence in accordance with $F_R$ shown in Fig.~\ref{FIG:ued_thz_measurements}~(c). Furthermore, we clearly observe that $\Delta\sigma_1^{\textup{slow}}$ achieves a maximum at a fluence of $\sim$20~mJ/cm$^2$ beyond which it decreases, consistent with the behaviour of the (200) and (220) diffraction peaks and $F_{\mathscr{M}}$ shown in Fig.~\ref{FIG:ued_thz_measurements}~(a) and (c). By analyzing the conductivity terms in an effective medium model, we find good agreement for $F_{\mathscr{M}}$ in the metallic limit~\cite{SM}. 

\begin{figure}
	\centering
	\includegraphics[width = 0.4\textwidth]{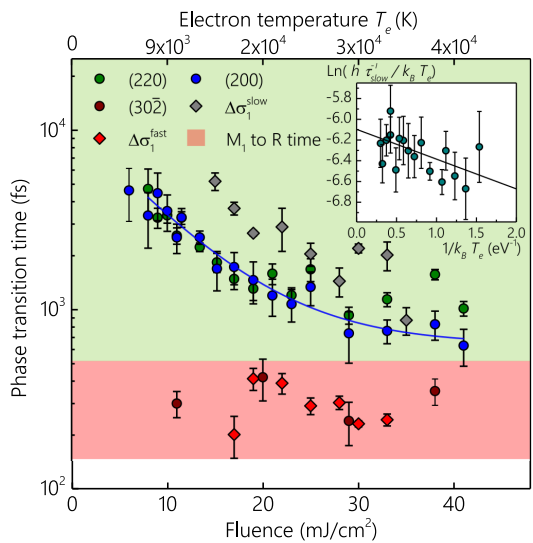}
    \caption{Time constants for the photoinduced phase transitions in VO$_2$. Time constants for the slow ((200) blue, (220) green) and fast (30$\bar{2}$, black) peak dynamics as determined from the UED data. THz time-domain spectroscopy measurements of the time constants pertaining to the slow and fast components of $\sigma_1$ are depicted by grey and red diamonds respectively. The red shaded area represents the temporal region ($350\pm~150~$fs) associated with the photo-induced structural phase transition from $M_1\rightarrow R$ which dominates at high pump fluence. The solid blue line is an exponential fit to the (200) peak time constants $\tau_{\textup{slow}}$. \textbf{Inset:} Plot of $\ln\left(h\tau_{\textup{slow}}^{-1}/k_B T_e\right)$ vs. inverse electron temperature $1/k_B T_e$ using values of $\tau_{\textup{slow}}$. The solid line is a fit to~\eqref{eqn:EyringPolanyi}, from which the activation energy $E_A$ and entropy $\Delta S^{\ddagger}$ are determined.}
    \label{FIG:time_consts}
\end{figure}

\subsection{Activation energy and kinetics}\label{sec:kinetics}
The $M_1\rightarrow R$ and $M_1\rightarrow \mathscr{M}$ transitions exhibit qualitatively different kinetic behaviour as evidenced by the fluence dependence of the time constants $\tau_{\textup{fast}}$ and $\tau_{\textup{slow}}$ obtained from both UED and TRTS (Fig.~\ref{FIG:time_consts}). The time constant for the $M_1\rightarrow R$ transition ($\tau_{\textup{fast}}$), captured by the dynamics of the (30$\bar{2}$) peak in the UED measurements and by $\tau_{\textup{fast}}$ in the THz measurements, is $350\pm100$ fs \textit{independent} of fluence. This demonstrates that photo-induced $M_1\rightarrow R$ transition -- the melting of the periodic lattice distortion -- is non-thermal and barrier free. The results for the $R$--phase volume fraction (Fig.~\ref{FIG:ued_thz_measurements}), however, also show that the excitation threshold for this non-thermal phase transition is heterogeneous in PLD grown films, depending on local crystallite size and strain conditions.  This last fact was also previously observed by Zewail using UEM~\cite{Lobastov2007} and others by nanoscopy~\cite{ocallahan2015,Donges2016}. The $M_1\rightarrow \mathscr{M}$ time constant, conversely, decreases significantly with pump fluence as seen in the (200) and (220) peak dynamics from UED and the $\tau_{\textup{slow}}$ from TRTS. The exponential increase in the $M_1\rightarrow \mathscr{M}$ rate with excitation energy deposited in the electron system strongly suggests that the $M_1\rightarrow \mathscr{M}$ is an activated process. We can extract the activation energy $E_A$ from this data by determining the electronic excitation energy $k_BT_e$ as a function of pump fluence~\cite{SM} and invoking the Eyring-Polanyi equation from transition state theory~\cite{EyringPolyani} 
\begin{equation}\label{eqn:EyringPolanyi}
\ln \left({\frac  {h \tau_{\textup{slow}}^{-1}}{k_B T_e}}\right)=-\frac{E_A}{k_B T_e}+\frac{\Delta S^{\ddagger }}{k_B},
\end{equation}
where $\Delta S^{\ddagger}$ is the entropy of activation. We take the values of $\tau_{\textup{slow}}$ for the (200) peak shown in Fig.~\ref{FIG:time_consts} and plot $\ln\left(h\tau_{\textup{slow}}^{-1}\big/k_B T_e\right)$ vs. $1\big/k_BT_e$ which is shown in the inset of Fig.~\ref{FIG:time_consts}. By fitting to Eqn.~\eqref{eqn:EyringPolanyi}, we determine $E_A=304\pm 109~$meV. This describes a fundamental property of the photo-induced $M_1\rightarrow\mathscr{M}$ transition. Furthermore, the fluence required to deposit $E_A$ per unit cell is $\mathscr{F}\approx 3.7~$mJ/cm$^2$, which is the value previously attributed to the $M_1\rightarrow R$ IMT threshold~\cite{Becker1994,Cavalleri2001,Kubler2007}. This is also in agreement with the fluence threshold extracted from the low fluence data points in Fig.~\ref{FIG:ued_thz_measurements}~(a). 


\section{Discussion and conclusion}\label{sec:discussion}
\begin{figure*}[t!]
	\centering
	\includegraphics[width = 0.85\textwidth]{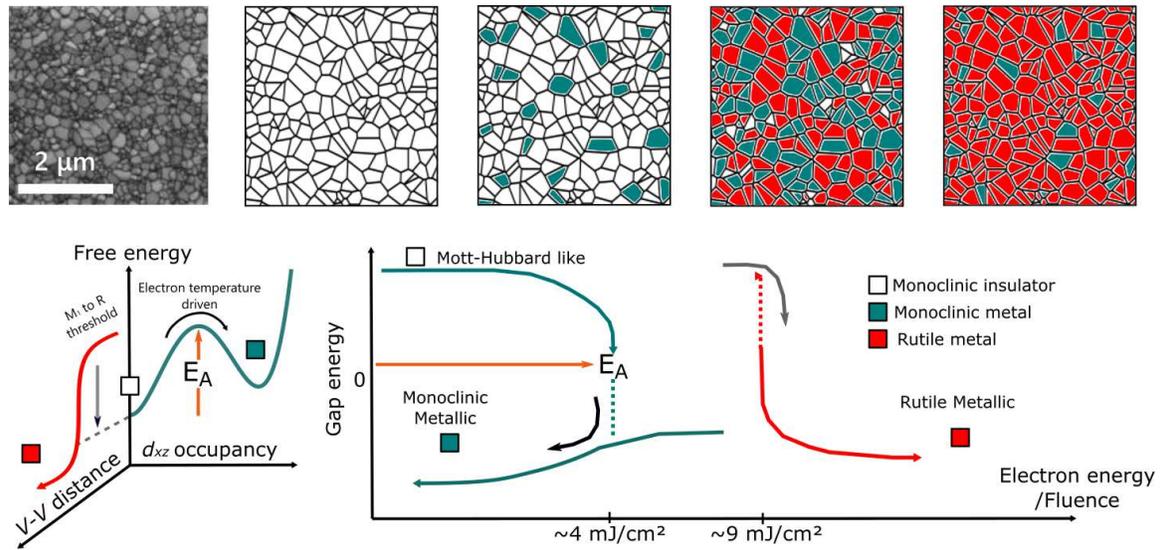}
    \caption{Illustration of the multi-phase heterogeneity of photo-excited VO$_2$. \textbf{Top: (Left to right)} Scanning electron microscope image of polycrystalline VO$_2$ grown by pulsed laser deposition. Graphical images of crystallites phases with increasing photoexcitation strength. Below the fluence threshold of $\sim 4~$mJ/cm$^2$ the response of the material is Mott-Hubbard like: instantaneous transient metallization from excited carriers followed by a recovery to the insulating phase within $\sim 100~$fs. When the fluence threshold for the $M_1\rightarrow \mathscr{M}$ IMT is reached in a particular crystallite, the gap collapses and the crystallite transitions to the $\mathscr{M}$ phase for $\sim 100~$ps. Eventually the crystallographic structural IMT fluence threshold ($\sim 9~$mJ/cm$^2$) is reached and crystallites transforms to the Rutile phase. \textbf{Bottom left:} Schematic representation of the free energy landscape. The monoclinic metal phase is described by a free energy minimum and the associated IMT is driven by electron temperature with an activation energy of $E_A=304\pm 109$ meV. The occupancy of the $d_{xz}$ orbital changes from partial to full filling during this process. The monoclinic-rutile structural IMT has a barrier which is removed when the threshold fluence of $\sim 9~$mJ/cm$^2$ is reached leading to an increase in the V-V distance (removal of dimerization).} 
    \label{FIG:summary}
\end{figure*}

We have demonstrated that photoexcitation of $M_1$ VO$_2$ yields a complex, heterogenous, multiphase film whose structure and properties are both time and fluence dependent. The character of the fluence depedent transformation is summarized in Fig.~\ref{FIG:summary}. At pump fluences below $\sim3$~mJ/cm$^2$ there is no long-lived ($>1~$ps) transformation of the $M_1$ structure, and VO$_2$ behaves like other Mott insulators insofar as optical excitation induces a relatively small, impulsive increase in conductivity followed by a complete recovery of the insulating state~\cite{Pashkin2011,Cocker2012,Morrison2014a,Wegkamp2014}. Above $\sim3$~mJ/cm$^2$, however, photoexcitation stimulates a phase transition in the electron system that stabilizes metallic properties through an orbitally selective charge re-organization: the $\mathscr{M}$ phase. Between 3-8~mJ/cm$^2$ photoexcitation exclusively yields the $\mathscr{M}$ phase, which populates ~15--20$\%$ of the film by 8~mJ/cm$^2$. In this fluence range, time-resolved photoemission experiments show a complete collapse of the bandgap~\cite{Wegkamp2014}, TRTS experiments show a dramatic increase in conductivity~\cite{Pashkin2011,Cocker2012} and optical studies show large changes in the dielectric function~\cite{Cavalleri2001,wall2012,Jager2017}, all of which are persistent, long-lived and characteristic of a phase-transition. Given the nature of the equilibrium phase diagram, these observations were previously interpreted as evidence of the $M_1\rightarrow R$ transition. The $M_1\rightarrow R$ transition, however, exhibits a minimum fluence threshold of $\sim$8--9~mJ/cm$^2$ consistent with surface sensitive experiments~\cite{Baum2007,Lobastov2007} and coherent phonon investigations~\cite{wall2012}. Above 8~mJ/cm$^2$ photoexcitation yields a heterogenous response with both $\mathscr{M}$ and $R$ phase fractions increasing with fluence to approximately 20~mJ/cm$^2$ where each phase occupies $\sim50\%$ of the film. At higher fluences $M_1\rightarrow R$ dominates and the $\mathscr{M}$ phase occupies a decreasing proportion of the film. 

Non-thermal melting as a route to the control of material structure and properties with femtosecond laser excitation has been known for some time and there are examples in several material classes.  Much more novel is the ($M_1\rightarrow \mathscr{M}$) which has no equilibrium analog and represents a new direction for using optical excitation to control the properties of strongly correlated materials. The $M_1\rightarrow \mathscr{M}$ is thermally activated and does not involve a significant lattice structural component, representing a phase transition in the electron system alone. Our results are consistent with the recent computations of He and Millis~\cite{Millis2016} which indicate that an orbital selective transition can be driven in $M_1$ VO$_2$ through the increase in electron temperature following femtosecond laser excitation. This transition depletes the occupancy of the V--$3d_{x^2-y^2}$ band that is split by the V--V dimerization in favour of the V--$3d_{xz}$ band that mixes strongly with the O--$2p$ orbitals due to the anti-ferroelectric tilting of the V--V dimers (Fig.~\ref{FIG:atomic_potential_maps}~(f)). The $M_1$ bandgap collapses along with this transition yielding a metallic phase. The three salient features of this picture are in agreement with our observatons; thermal activation on the order of $100~$meV, orbital selection and bandgap collapse to a metallic phase. Of interest is the fact that depletion of the V--$3d_{x^2-y^2}$ band where states are expected to be localized on the V--V dimers does not seem to significantly lengthen the V--V dimer bond. The question arises whether this phenomena can be entirely understood inside a picture that treats $M_1$ VO$_2$ as a $d^1$ system, or whether more than a single V-$3d$ electron is involved as DMFT calculations suggest~\cite{Weber2012}. Such DMFT results from Weber~\cite{Weber2012} suggest that $M_1$ VO$_2$ is a paramagnetic metal with antiferroelectric character, like that shown in Fig.~\ref{FIG:atomic_potential_maps}~(d), when intra-dimer correlations are not included.    

The combination of UED and TRTS measurements also makes it possible to address the question of a structural bottleneck associated with the photoinduced IMT in $M_1$ VO$_2$.  Here we definitively show that the timescale associated with the IMT, i.e. the timescale associated with the emergence of metallic conductivity similar to that of the equilibrium metallic phase, is determined by that of the structural phase transitions (Fig.~\ref{FIG:ued_thz_time_traces} and~\ref{FIG:time_consts}. Following photoexcitation at sufficient fluence, there is overwhelming evidence that there is an impulsive collapse of the bandgap in $M_1$ VO$_2$ with equally rapid changes in optical properties. However, the emergence of metallic transport properties occurs on the same timescale as the structural phase transitions.  Clearly the localization-delocalization transition that leads to a 5 order of magnitude increase  in conductivty is inseparable from the structural phase transitions.           

In conclusion, we have combined UED and TRTS measurements of VO$_2$ and decoupled the concurrent structural phase transitions along with their contribution to the multiphase heterogeneity of the sample following photoexcitation.  We have shown that the monoclinc metal phase is the product of a thermally activated transition in the electron system, which provides a new avenue for the optical control of strongly correlated material properties.  

\bibliographystyle{apsrev}
\bibliography{Mendeley}

\begin{thebibliography}{45}
\expandafter\ifx\csname natexlab\endcsname\relax\def\natexlab#1{#1}\fi
\expandafter\ifx\csname bibnamefont\endcsname\relax
  \def\bibnamefont#1{#1}\fi
\expandafter\ifx\csname bibfnamefont\endcsname\relax
  \def\bibfnamefont#1{#1}\fi
\expandafter\ifx\csname citenamefont\endcsname\relax
  \def\citenamefont#1{#1}\fi
\expandafter\ifx\csname url\endcsname\relax
  \def\url#1{\texttt{#1}}\fi
\expandafter\ifx\csname urlprefix\endcsname\relax\def\urlprefix{URL }\fi
\providecommand{\bibinfo}[2]{#2}
\providecommand{\eprint}[2][]{\url{#2}}

\bibitem[{\citenamefont{Morin}(1959)}]{Morin1959}
\bibinfo{author}{\bibfnamefont{F.~J.} \bibnamefont{Morin}},
  \bibinfo{journal}{Phys. Rev. Lett.} \textbf{\bibinfo{volume}{3}},
  \bibinfo{pages}{34} (\bibinfo{year}{1959}),
  \urlprefix\url{https://link.aps.org/doi/10.1103/PhysRevLett.3.34}.

\bibitem[{\citenamefont{Goodenough}(1971)}]{GOODENOUGH1971}
\bibinfo{author}{\bibfnamefont{J.~B.} \bibnamefont{Goodenough}},
  \bibinfo{journal}{Journal of Solid State Chemistry}
  \textbf{\bibinfo{volume}{3}}, \bibinfo{pages}{490 } (\bibinfo{year}{1971}),
  ISSN \bibinfo{issn}{0022-4596},
  \urlprefix\url{http://www.sciencedirect.com/science/article/pii/0022459671900910}.

\bibitem[{\citenamefont{Zylbersztejn and Mott}(1975)}]{Zylbersztejn1975}
\bibinfo{author}{\bibfnamefont{A.}~\bibnamefont{Zylbersztejn}}
  \bibnamefont{and} \bibinfo{author}{\bibfnamefont{N.~F.} \bibnamefont{Mott}},
  \bibinfo{journal}{Phys. Rev. B} \textbf{\bibinfo{volume}{11}},
  \bibinfo{pages}{4383} (\bibinfo{year}{1975}),
  \urlprefix\url{https://link.aps.org/doi/10.1103/PhysRevB.11.4383}.

\bibitem[{\citenamefont{Eyert}(2002)}]{Eyert2002}
\bibinfo{author}{\bibfnamefont{V.}~\bibnamefont{Eyert}},
  \bibinfo{journal}{Annalen der Physik} \textbf{\bibinfo{volume}{11}},
  \bibinfo{pages}{650} (\bibinfo{year}{2002}), ISSN \bibinfo{issn}{1521-3889},
  \urlprefix\url{http://dx.doi.org/10.1002/1521-3889(200210)11:9<650::AID-ANDP650>3.0.CO;2-K}.

\bibitem[{\citenamefont{Basov et~al.}(2011)\citenamefont{Basov, Averitt,
  van~der Marel, Dressel, and Haule}}]{BasovRMP2011}
\bibinfo{author}{\bibfnamefont{D.~N.} \bibnamefont{Basov}},
  \bibinfo{author}{\bibfnamefont{R.~D.} \bibnamefont{Averitt}},
  \bibinfo{author}{\bibfnamefont{D.}~\bibnamefont{van~der Marel}},
  \bibinfo{author}{\bibfnamefont{M.}~\bibnamefont{Dressel}}, \bibnamefont{and}
  \bibinfo{author}{\bibfnamefont{K.}~\bibnamefont{Haule}},
  \bibinfo{journal}{Rev. Mod. Phys.} \textbf{\bibinfo{volume}{83}},
  \bibinfo{pages}{471} (\bibinfo{year}{2011}),
  \urlprefix\url{https://link.aps.org/doi/10.1103/RevModPhys.83.471}.

\bibitem[{\citenamefont{Budai et~al.}(2014)\citenamefont{Budai, Hong, Manley,
  Specht, Li, Tischler, Abernathy, Said, Leu, Boatner et~al.}}]{budai2014}
\bibinfo{author}{\bibfnamefont{J.~D.} \bibnamefont{Budai}},
  \bibinfo{author}{\bibfnamefont{J.}~\bibnamefont{Hong}},
  \bibinfo{author}{\bibfnamefont{M.~E.} \bibnamefont{Manley}},
  \bibinfo{author}{\bibfnamefont{E.~D.} \bibnamefont{Specht}},
  \bibinfo{author}{\bibfnamefont{C.~W.} \bibnamefont{Li}},
  \bibinfo{author}{\bibfnamefont{J.~Z.} \bibnamefont{Tischler}},
  \bibinfo{author}{\bibfnamefont{D.~L.} \bibnamefont{Abernathy}},
  \bibinfo{author}{\bibfnamefont{A.~H.} \bibnamefont{Said}},
  \bibinfo{author}{\bibfnamefont{B.~M.} \bibnamefont{Leu}},
  \bibinfo{author}{\bibfnamefont{L.~A.} \bibnamefont{Boatner}},
  \bibnamefont{et~al.}, \bibinfo{journal}{Nature}
  \textbf{\bibinfo{volume}{515}}, \bibinfo{pages}{535} (\bibinfo{year}{2014}).

\bibitem[{\citenamefont{Hendaoui et~al.}(2013)\citenamefont{Hendaoui,
  \'{E}mond, Chaker, and \'{E}mile Haddad}}]{Hendaoui2013}
\bibinfo{author}{\bibfnamefont{A.}~\bibnamefont{Hendaoui}},
  \bibinfo{author}{\bibfnamefont{N.}~\bibnamefont{\'{E}mond}},
  \bibinfo{author}{\bibfnamefont{M.}~\bibnamefont{Chaker}}, \bibnamefont{and}
  \bibinfo{author}{\bibnamefont{\'{E}mile Haddad}}, \bibinfo{journal}{Applied
  Physics Letters} \textbf{\bibinfo{volume}{102}}, \bibinfo{pages}{061107}
  (\bibinfo{year}{2013}), \urlprefix\url{https://doi.org/10.1063/1.4792277}.

\bibitem[{\citenamefont{Ryckman et~al.}(2012)\citenamefont{Ryckman,
  Diez-Blanco, Nag, Marvel, Choi, Haglund, and Weiss}}]{Ryckman2012}
\bibinfo{author}{\bibfnamefont{J.~D.} \bibnamefont{Ryckman}},
  \bibinfo{author}{\bibfnamefont{V.}~\bibnamefont{Diez-Blanco}},
  \bibinfo{author}{\bibfnamefont{J.}~\bibnamefont{Nag}},
  \bibinfo{author}{\bibfnamefont{R.~E.} \bibnamefont{Marvel}},
  \bibinfo{author}{\bibfnamefont{B.~K.} \bibnamefont{Choi}},
  \bibinfo{author}{\bibfnamefont{R.~F.} \bibnamefont{Haglund}},
  \bibnamefont{and} \bibinfo{author}{\bibfnamefont{S.~M.} \bibnamefont{Weiss}},
  \bibinfo{journal}{Opt. Express} \textbf{\bibinfo{volume}{20}},
  \bibinfo{pages}{13215} (\bibinfo{year}{2012}),
  \urlprefix\url{http://www.opticsexpress.org/abstract.cfm?URI=oe-20-12-13215}.

\bibitem[{\citenamefont{Miller et~al.}(2017)\citenamefont{Miller, Hallman,
  Haglund, and Weiss}}]{Miller2017}
\bibinfo{author}{\bibfnamefont{K.~J.} \bibnamefont{Miller}},
  \bibinfo{author}{\bibfnamefont{K.~A.} \bibnamefont{Hallman}},
  \bibinfo{author}{\bibfnamefont{R.~F.} \bibnamefont{Haglund}},
  \bibnamefont{and} \bibinfo{author}{\bibfnamefont{S.~M.} \bibnamefont{Weiss}},
  \bibinfo{journal}{Opt. Express} \textbf{\bibinfo{volume}{25}},
  \bibinfo{pages}{26527} (\bibinfo{year}{2017}),
  \urlprefix\url{http://www.opticsexpress.org/abstract.cfm?URI=oe-25-22-26527}.

\bibitem[{\citenamefont{Mott and Friedman}(1974)}]{Mott1974}
\bibinfo{author}{\bibfnamefont{N.~F.} \bibnamefont{Mott}} \bibnamefont{and}
  \bibinfo{author}{\bibfnamefont{L.}~\bibnamefont{Friedman}},
  \bibinfo{journal}{The Philosophical Magazine: A Journal of Theoretical
  Experimental and Applied Physics} \textbf{\bibinfo{volume}{30}},
  \bibinfo{pages}{389} (\bibinfo{year}{1974}).

\bibitem[{\citenamefont{Eyert}(2011)}]{Eyert2011}
\bibinfo{author}{\bibfnamefont{V.}~\bibnamefont{Eyert}},
  \bibinfo{journal}{Phys. Rev. Lett.} \textbf{\bibinfo{volume}{107}},
  \bibinfo{pages}{016401} (\bibinfo{year}{2011}),
  \urlprefix\url{https://link.aps.org/doi/10.1103/PhysRevLett.107.016401}.

\bibitem[{\citenamefont{Weber et~al.}(2012)\citenamefont{Weber, O'Regan, Hine,
  Payne, Kotliar, and Littlewood}}]{Weber2012}
\bibinfo{author}{\bibfnamefont{C.}~\bibnamefont{Weber}},
  \bibinfo{author}{\bibfnamefont{D.~D.} \bibnamefont{O'Regan}},
  \bibinfo{author}{\bibfnamefont{N.~D.~M.} \bibnamefont{Hine}},
  \bibinfo{author}{\bibfnamefont{M.~C.} \bibnamefont{Payne}},
  \bibinfo{author}{\bibfnamefont{G.}~\bibnamefont{Kotliar}}, \bibnamefont{and}
  \bibinfo{author}{\bibfnamefont{P.~B.} \bibnamefont{Littlewood}},
  \bibinfo{journal}{Phys. Rev. Lett.} \textbf{\bibinfo{volume}{108}},
  \bibinfo{pages}{256402} (\bibinfo{year}{2012}),
  \urlprefix\url{https://link.aps.org/doi/10.1103/PhysRevLett.108.256402}.

\bibitem[{\citenamefont{Biermann et~al.}(2005)\citenamefont{Biermann,
  Poteryaev, Lichtenstein, and Georges}}]{Biermann2005}
\bibinfo{author}{\bibfnamefont{S.}~\bibnamefont{Biermann}},
  \bibinfo{author}{\bibfnamefont{A.}~\bibnamefont{Poteryaev}},
  \bibinfo{author}{\bibfnamefont{A.~I.} \bibnamefont{Lichtenstein}},
  \bibnamefont{and} \bibinfo{author}{\bibfnamefont{A.}~\bibnamefont{Georges}},
  \bibinfo{journal}{Phys. Rev. Lett.} \textbf{\bibinfo{volume}{94}},
  \bibinfo{pages}{026404} (\bibinfo{year}{2005}),
  \urlprefix\url{https://link.aps.org/doi/10.1103/PhysRevLett.94.026404}.

\bibitem[{\citenamefont{Brito et~al.}(2016)\citenamefont{Brito, Aguiar, Haule,
  and Kotliar}}]{Brito2016}
\bibinfo{author}{\bibfnamefont{W.~H.} \bibnamefont{Brito}},
  \bibinfo{author}{\bibfnamefont{M.~C.~O.} \bibnamefont{Aguiar}},
  \bibinfo{author}{\bibfnamefont{K.}~\bibnamefont{Haule}}, \bibnamefont{and}
  \bibinfo{author}{\bibfnamefont{G.}~\bibnamefont{Kotliar}},
  \bibinfo{journal}{Phys. Rev. Lett.} \textbf{\bibinfo{volume}{117}},
  \bibinfo{pages}{056402} (\bibinfo{year}{2016}),
  \urlprefix\url{https://link.aps.org/doi/10.1103/PhysRevLett.117.056402}.

\bibitem[{\citenamefont{Huffman et~al.}(2017)\citenamefont{Huffman, Hendriks,
  Walter, Yoon, Ju, Smith, Carr, Krakauer, and Qazilbash}}]{Huffman2017}
\bibinfo{author}{\bibfnamefont{T.~J.} \bibnamefont{Huffman}},
  \bibinfo{author}{\bibfnamefont{C.}~\bibnamefont{Hendriks}},
  \bibinfo{author}{\bibfnamefont{E.~J.} \bibnamefont{Walter}},
  \bibinfo{author}{\bibfnamefont{J.}~\bibnamefont{Yoon}},
  \bibinfo{author}{\bibfnamefont{H.}~\bibnamefont{Ju}},
  \bibinfo{author}{\bibfnamefont{R.}~\bibnamefont{Smith}},
  \bibinfo{author}{\bibfnamefont{G.~L.} \bibnamefont{Carr}},
  \bibinfo{author}{\bibfnamefont{H.}~\bibnamefont{Krakauer}}, \bibnamefont{and}
  \bibinfo{author}{\bibfnamefont{M.~M.} \bibnamefont{Qazilbash}},
  \bibinfo{journal}{Phys. Rev. B} \textbf{\bibinfo{volume}{95}},
  \bibinfo{pages}{075125} (\bibinfo{year}{2017}),
  \urlprefix\url{https://link.aps.org/doi/10.1103/PhysRevB.95.075125}.

\bibitem[{\citenamefont{N\'ajera et~al.}(2018)\citenamefont{N\'ajera, Civelli,
  Dobrosavljevi\ifmmode~\acute{c}\else \'{c}\fi{}, and Rozenberg}}]{Najera2018}
\bibinfo{author}{\bibfnamefont{O.}~\bibnamefont{N\'ajera}},
  \bibinfo{author}{\bibfnamefont{M.}~\bibnamefont{Civelli}},
  \bibinfo{author}{\bibfnamefont{V.}~\bibnamefont{Dobrosavljevi\ifmmode~\acute{c}\else
  \'{c}\fi{}}}, \bibnamefont{and} \bibinfo{author}{\bibfnamefont{M.~J.}
  \bibnamefont{Rozenberg}}, \bibinfo{journal}{Phys. Rev. B}
  \textbf{\bibinfo{volume}{97}}, \bibinfo{pages}{045108}
  (\bibinfo{year}{2018}),
  \urlprefix\url{https://link.aps.org/doi/10.1103/PhysRevB.97.045108}.

\bibitem[{\citenamefont{Becker et~al.}(1994)\citenamefont{Becker, Buckman,
  Walser, Lépine, Georges, and Brun}}]{Becker1994}
\bibinfo{author}{\bibfnamefont{M.~F.} \bibnamefont{Becker}},
  \bibinfo{author}{\bibfnamefont{A.~B.} \bibnamefont{Buckman}},
  \bibinfo{author}{\bibfnamefont{R.~M.} \bibnamefont{Walser}},
  \bibinfo{author}{\bibfnamefont{T.}~\bibnamefont{Lépine}},
  \bibinfo{author}{\bibfnamefont{P.}~\bibnamefont{Georges}}, \bibnamefont{and}
  \bibinfo{author}{\bibfnamefont{A.}~\bibnamefont{Brun}},
  \bibinfo{journal}{Applied Physics Letters} \textbf{\bibinfo{volume}{65}},
  \bibinfo{pages}{1507} (\bibinfo{year}{1994}),
  \urlprefix\url{https://doi.org/10.1063/1.112974}.

\bibitem[{\citenamefont{Cavalleri et~al.}(2001)\citenamefont{Cavalleri, T\'oth,
  Siders, Squier, R\'aksi, Forget, and Kieffer}}]{Cavalleri2001}
\bibinfo{author}{\bibfnamefont{A.}~\bibnamefont{Cavalleri}},
  \bibinfo{author}{\bibfnamefont{C.}~\bibnamefont{T\'oth}},
  \bibinfo{author}{\bibfnamefont{C.~W.} \bibnamefont{Siders}},
  \bibinfo{author}{\bibfnamefont{J.~A.} \bibnamefont{Squier}},
  \bibinfo{author}{\bibfnamefont{F.}~\bibnamefont{R\'aksi}},
  \bibinfo{author}{\bibfnamefont{P.}~\bibnamefont{Forget}}, \bibnamefont{and}
  \bibinfo{author}{\bibfnamefont{J.~C.} \bibnamefont{Kieffer}},
  \bibinfo{journal}{Phys. Rev. Lett.} \textbf{\bibinfo{volume}{87}},
  \bibinfo{pages}{237401} (\bibinfo{year}{2001}),
  \urlprefix\url{https://link.aps.org/doi/10.1103/PhysRevLett.87.237401}.

\bibitem[{\citenamefont{Hada et~al.}(2011)\citenamefont{Hada, Okimura, and
  Matsuo}}]{Hada2011}
\bibinfo{author}{\bibfnamefont{M.}~\bibnamefont{Hada}},
  \bibinfo{author}{\bibfnamefont{K.}~\bibnamefont{Okimura}}, \bibnamefont{and}
  \bibinfo{author}{\bibfnamefont{J.}~\bibnamefont{Matsuo}},
  \bibinfo{journal}{Applied Physics Letters} \textbf{\bibinfo{volume}{99}},
  \bibinfo{pages}{051903} (\bibinfo{year}{2011}),
  \urlprefix\url{https://doi.org/10.1063/1.3621900}.

\bibitem[{\citenamefont{Baum et~al.}(2007)\citenamefont{Baum, Yang, and
  Zewail}}]{Baum2007}
\bibinfo{author}{\bibfnamefont{P.}~\bibnamefont{Baum}},
  \bibinfo{author}{\bibfnamefont{D.-S.} \bibnamefont{Yang}}, \bibnamefont{and}
  \bibinfo{author}{\bibfnamefont{A.~H.} \bibnamefont{Zewail}},
  \bibinfo{journal}{Science} \textbf{\bibinfo{volume}{318}},
  \bibinfo{pages}{788} (\bibinfo{year}{2007}), ISSN \bibinfo{issn}{0036-8075},
  \urlprefix\url{http://science.sciencemag.org/content/318/5851/788}.

\bibitem[{\citenamefont{Morrison et~al.}(2014)\citenamefont{Morrison,
  Chatelain, Tiwari, Hendaoui, Bruhacs, Chaker, and Siwick}}]{Morrison2014a}
\bibinfo{author}{\bibfnamefont{V.~R.} \bibnamefont{Morrison}},
  \bibinfo{author}{\bibfnamefont{R.~P.} \bibnamefont{Chatelain}},
  \bibinfo{author}{\bibfnamefont{K.~L.} \bibnamefont{Tiwari}},
  \bibinfo{author}{\bibfnamefont{A.}~\bibnamefont{Hendaoui}},
  \bibinfo{author}{\bibfnamefont{A.}~\bibnamefont{Bruhacs}},
  \bibinfo{author}{\bibfnamefont{M.}~\bibnamefont{Chaker}}, \bibnamefont{and}
  \bibinfo{author}{\bibfnamefont{B.~J.} \bibnamefont{Siwick}},
  \bibinfo{journal}{Science} \textbf{\bibinfo{volume}{346}},
  \bibinfo{pages}{445} (\bibinfo{year}{2014}), ISSN \bibinfo{issn}{0036-8075}.

\bibitem[{\citenamefont{Tao et~al.}(2016)\citenamefont{Tao, Zhou, Han, Torres,
  Wang, Sepulveda, Chang, Young, Lunt, and Ruan}}]{tao2016}
\bibinfo{author}{\bibfnamefont{Z.}~\bibnamefont{Tao}},
  \bibinfo{author}{\bibfnamefont{F.}~\bibnamefont{Zhou}},
  \bibinfo{author}{\bibfnamefont{T.-R.~T.} \bibnamefont{Han}},
  \bibinfo{author}{\bibfnamefont{D.}~\bibnamefont{Torres}},
  \bibinfo{author}{\bibfnamefont{T.}~\bibnamefont{Wang}},
  \bibinfo{author}{\bibfnamefont{N.}~\bibnamefont{Sepulveda}},
  \bibinfo{author}{\bibfnamefont{K.}~\bibnamefont{Chang}},
  \bibinfo{author}{\bibfnamefont{M.}~\bibnamefont{Young}},
  \bibinfo{author}{\bibfnamefont{R.~R.} \bibnamefont{Lunt}}, \bibnamefont{and}
  \bibinfo{author}{\bibfnamefont{C.-Y.} \bibnamefont{Ruan}},
  \bibinfo{journal}{Scientific reports} \textbf{\bibinfo{volume}{6}},
  \bibinfo{pages}{38514} (\bibinfo{year}{2016}).

\bibitem[{\citenamefont{Cavalleri et~al.}(2005)\citenamefont{Cavalleri, Rini,
  Chong, Fourmaux, Glover, Heimann, Kieffer, and Schoenlein}}]{Cavalleri2005}
\bibinfo{author}{\bibfnamefont{A.}~\bibnamefont{Cavalleri}},
  \bibinfo{author}{\bibfnamefont{M.}~\bibnamefont{Rini}},
  \bibinfo{author}{\bibfnamefont{H.~H.~W.} \bibnamefont{Chong}},
  \bibinfo{author}{\bibfnamefont{S.}~\bibnamefont{Fourmaux}},
  \bibinfo{author}{\bibfnamefont{T.~E.} \bibnamefont{Glover}},
  \bibinfo{author}{\bibfnamefont{P.~A.} \bibnamefont{Heimann}},
  \bibinfo{author}{\bibfnamefont{J.~C.} \bibnamefont{Kieffer}},
  \bibnamefont{and} \bibinfo{author}{\bibfnamefont{R.~W.}
  \bibnamefont{Schoenlein}}, \bibinfo{journal}{Phys. Rev. Lett.}
  \textbf{\bibinfo{volume}{95}}, \bibinfo{pages}{067405}
  (\bibinfo{year}{2005}),
  \urlprefix\url{https://link.aps.org/doi/10.1103/PhysRevLett.95.067405}.

\bibitem[{\citenamefont{Haverkort et~al.}(2005)\citenamefont{Haverkort, Hu,
  Tanaka, Reichelt, Streltsov, Korotin, Anisimov, Hsieh, Lin, Chen
  et~al.}}]{Haverkort2005}
\bibinfo{author}{\bibfnamefont{M.~W.} \bibnamefont{Haverkort}},
  \bibinfo{author}{\bibfnamefont{Z.}~\bibnamefont{Hu}},
  \bibinfo{author}{\bibfnamefont{A.}~\bibnamefont{Tanaka}},
  \bibinfo{author}{\bibfnamefont{W.}~\bibnamefont{Reichelt}},
  \bibinfo{author}{\bibfnamefont{S.~V.} \bibnamefont{Streltsov}},
  \bibinfo{author}{\bibfnamefont{M.~A.} \bibnamefont{Korotin}},
  \bibinfo{author}{\bibfnamefont{V.~I.} \bibnamefont{Anisimov}},
  \bibinfo{author}{\bibfnamefont{H.~H.} \bibnamefont{Hsieh}},
  \bibinfo{author}{\bibfnamefont{H.-J.} \bibnamefont{Lin}},
  \bibinfo{author}{\bibfnamefont{C.~T.} \bibnamefont{Chen}},
  \bibnamefont{et~al.}, \bibinfo{journal}{Phys. Rev. Lett.}
  \textbf{\bibinfo{volume}{95}}, \bibinfo{pages}{196404}
  (\bibinfo{year}{2005}),
  \urlprefix\url{https://link.aps.org/doi/10.1103/PhysRevLett.95.196404}.

\bibitem[{\citenamefont{Wegkamp et~al.}(2014)\citenamefont{Wegkamp, Herzog,
  Xian, Gatti, Cudazzo, McGahan, Marvel, Haglund, Rubio, Wolf
  et~al.}}]{Wegkamp2014}
\bibinfo{author}{\bibfnamefont{D.}~\bibnamefont{Wegkamp}},
  \bibinfo{author}{\bibfnamefont{M.}~\bibnamefont{Herzog}},
  \bibinfo{author}{\bibfnamefont{L.}~\bibnamefont{Xian}},
  \bibinfo{author}{\bibfnamefont{M.}~\bibnamefont{Gatti}},
  \bibinfo{author}{\bibfnamefont{P.}~\bibnamefont{Cudazzo}},
  \bibinfo{author}{\bibfnamefont{C.~L.} \bibnamefont{McGahan}},
  \bibinfo{author}{\bibfnamefont{R.~E.} \bibnamefont{Marvel}},
  \bibinfo{author}{\bibfnamefont{R.~F.} \bibnamefont{Haglund}},
  \bibinfo{author}{\bibfnamefont{A.}~\bibnamefont{Rubio}},
  \bibinfo{author}{\bibfnamefont{M.}~\bibnamefont{Wolf}}, \bibnamefont{et~al.},
  \bibinfo{journal}{Phys. Rev. Lett.} \textbf{\bibinfo{volume}{113}},
  \bibinfo{pages}{216401} (\bibinfo{year}{2014}),
  \urlprefix\url{https://link.aps.org/doi/10.1103/PhysRevLett.113.216401}.

\bibitem[{\citenamefont{Cavalleri et~al.}(2004)\citenamefont{Cavalleri,
  Dekorsy, Chong, Kieffer, and Schoenlein}}]{Cavalleri2004}
\bibinfo{author}{\bibfnamefont{A.}~\bibnamefont{Cavalleri}},
  \bibinfo{author}{\bibfnamefont{T.}~\bibnamefont{Dekorsy}},
  \bibinfo{author}{\bibfnamefont{H.~H.~W.} \bibnamefont{Chong}},
  \bibinfo{author}{\bibfnamefont{J.~C.} \bibnamefont{Kieffer}},
  \bibnamefont{and} \bibinfo{author}{\bibfnamefont{R.~W.}
  \bibnamefont{Schoenlein}}, \bibinfo{journal}{Phys. Rev. B}
  \textbf{\bibinfo{volume}{70}}, \bibinfo{pages}{161102}
  (\bibinfo{year}{2004}),
  \urlprefix\url{https://link.aps.org/doi/10.1103/PhysRevB.70.161102}.

\bibitem[{\citenamefont{K\"ubler et~al.}(2007)\citenamefont{K\"ubler, Ehrke,
  Huber, Lopez, Halabica, Haglund, and Leitenstorfer}}]{Kubler2007}
\bibinfo{author}{\bibfnamefont{C.}~\bibnamefont{K\"ubler}},
  \bibinfo{author}{\bibfnamefont{H.}~\bibnamefont{Ehrke}},
  \bibinfo{author}{\bibfnamefont{R.}~\bibnamefont{Huber}},
  \bibinfo{author}{\bibfnamefont{R.}~\bibnamefont{Lopez}},
  \bibinfo{author}{\bibfnamefont{A.}~\bibnamefont{Halabica}},
  \bibinfo{author}{\bibfnamefont{R.~F.} \bibnamefont{Haglund}},
  \bibnamefont{and}
  \bibinfo{author}{\bibfnamefont{A.}~\bibnamefont{Leitenstorfer}},
  \bibinfo{journal}{Phys. Rev. Lett.} \textbf{\bibinfo{volume}{99}},
  \bibinfo{pages}{116401} (\bibinfo{year}{2007}),
  \urlprefix\url{https://link.aps.org/doi/10.1103/PhysRevLett.99.116401}.

\bibitem[{\citenamefont{Pashkin et~al.}(2011)\citenamefont{Pashkin, K\"ubler,
  Ehrke, Lopez, Halabica, Haglund, Huber, and Leitenstorfer}}]{Pashkin2011}
\bibinfo{author}{\bibfnamefont{A.}~\bibnamefont{Pashkin}},
  \bibinfo{author}{\bibfnamefont{C.}~\bibnamefont{K\"ubler}},
  \bibinfo{author}{\bibfnamefont{H.}~\bibnamefont{Ehrke}},
  \bibinfo{author}{\bibfnamefont{R.}~\bibnamefont{Lopez}},
  \bibinfo{author}{\bibfnamefont{A.}~\bibnamefont{Halabica}},
  \bibinfo{author}{\bibfnamefont{R.~F.} \bibnamefont{Haglund}},
  \bibinfo{author}{\bibfnamefont{R.}~\bibnamefont{Huber}}, \bibnamefont{and}
  \bibinfo{author}{\bibfnamefont{A.}~\bibnamefont{Leitenstorfer}},
  \bibinfo{journal}{Phys. Rev. B} \textbf{\bibinfo{volume}{83}},
  \bibinfo{pages}{195120} (\bibinfo{year}{2011}),
  \urlprefix\url{https://link.aps.org/doi/10.1103/PhysRevB.83.195120}.

\bibitem[{\citenamefont{Wall et~al.}(2012)\citenamefont{Wall, Wegkamp, Foglia,
  Appavoo, Nag, Haglund~Jr, St{\"a}hler, and Wolf}}]{wall2012}
\bibinfo{author}{\bibfnamefont{S.}~\bibnamefont{Wall}},
  \bibinfo{author}{\bibfnamefont{D.}~\bibnamefont{Wegkamp}},
  \bibinfo{author}{\bibfnamefont{L.}~\bibnamefont{Foglia}},
  \bibinfo{author}{\bibfnamefont{K.}~\bibnamefont{Appavoo}},
  \bibinfo{author}{\bibfnamefont{J.}~\bibnamefont{Nag}},
  \bibinfo{author}{\bibfnamefont{R.}~\bibnamefont{Haglund~Jr}},
  \bibinfo{author}{\bibfnamefont{J.}~\bibnamefont{St{\"a}hler}},
  \bibnamefont{and} \bibinfo{author}{\bibfnamefont{M.}~\bibnamefont{Wolf}},
  \bibinfo{journal}{Nature Communications} \textbf{\bibinfo{volume}{3}},
  \bibinfo{pages}{721} (\bibinfo{year}{2012}).

\bibitem[{\citenamefont{Cocker et~al.}(2012)\citenamefont{Cocker, Titova,
  Fourmaux, Holloway, Bandulet, Brassard, Kieffer, El~Khakani, and
  Hegmann}}]{Cocker2012}
\bibinfo{author}{\bibfnamefont{T.~L.} \bibnamefont{Cocker}},
  \bibinfo{author}{\bibfnamefont{L.~V.} \bibnamefont{Titova}},
  \bibinfo{author}{\bibfnamefont{S.}~\bibnamefont{Fourmaux}},
  \bibinfo{author}{\bibfnamefont{G.}~\bibnamefont{Holloway}},
  \bibinfo{author}{\bibfnamefont{H.-C.} \bibnamefont{Bandulet}},
  \bibinfo{author}{\bibfnamefont{D.}~\bibnamefont{Brassard}},
  \bibinfo{author}{\bibfnamefont{J.-C.} \bibnamefont{Kieffer}},
  \bibinfo{author}{\bibfnamefont{M.~A.} \bibnamefont{El~Khakani}},
  \bibnamefont{and} \bibinfo{author}{\bibfnamefont{F.~A.}
  \bibnamefont{Hegmann}}, \bibinfo{journal}{Phys. Rev. B}
  \textbf{\bibinfo{volume}{85}}, \bibinfo{pages}{155120}
  (\bibinfo{year}{2012}),
  \urlprefix\url{https://link.aps.org/doi/10.1103/PhysRevB.85.155120}.

\bibitem[{\citenamefont{Wall et~al.}(2013)\citenamefont{Wall, Foglia, Wegkamp,
  Appavoo, Nag, Haglund, St\"ahler, and Wolf}}]{Wall2013}
\bibinfo{author}{\bibfnamefont{S.}~\bibnamefont{Wall}},
  \bibinfo{author}{\bibfnamefont{L.}~\bibnamefont{Foglia}},
  \bibinfo{author}{\bibfnamefont{D.}~\bibnamefont{Wegkamp}},
  \bibinfo{author}{\bibfnamefont{K.}~\bibnamefont{Appavoo}},
  \bibinfo{author}{\bibfnamefont{J.}~\bibnamefont{Nag}},
  \bibinfo{author}{\bibfnamefont{R.~F.} \bibnamefont{Haglund}},
  \bibinfo{author}{\bibfnamefont{J.}~\bibnamefont{St\"ahler}},
  \bibnamefont{and} \bibinfo{author}{\bibfnamefont{M.}~\bibnamefont{Wolf}},
  \bibinfo{journal}{Phys. Rev. B} \textbf{\bibinfo{volume}{87}},
  \bibinfo{pages}{115126} (\bibinfo{year}{2013}),
  \urlprefix\url{https://link.aps.org/doi/10.1103/PhysRevB.87.115126}.

\bibitem[{\citenamefont{Mayer et~al.}(2015)\citenamefont{Mayer, Schmidt, Grupp,
  B\"uhler, Oelmann, Marvel, Haglund, Oka, Brida, Leitenstorfer
  et~al.}}]{Mayer2015}
\bibinfo{author}{\bibfnamefont{B.}~\bibnamefont{Mayer}},
  \bibinfo{author}{\bibfnamefont{C.}~\bibnamefont{Schmidt}},
  \bibinfo{author}{\bibfnamefont{A.}~\bibnamefont{Grupp}},
  \bibinfo{author}{\bibfnamefont{J.}~\bibnamefont{B\"uhler}},
  \bibinfo{author}{\bibfnamefont{J.}~\bibnamefont{Oelmann}},
  \bibinfo{author}{\bibfnamefont{R.~E.} \bibnamefont{Marvel}},
  \bibinfo{author}{\bibfnamefont{R.~F.} \bibnamefont{Haglund}},
  \bibinfo{author}{\bibfnamefont{T.}~\bibnamefont{Oka}},
  \bibinfo{author}{\bibfnamefont{D.}~\bibnamefont{Brida}},
  \bibinfo{author}{\bibfnamefont{A.}~\bibnamefont{Leitenstorfer}},
  \bibnamefont{et~al.}, \bibinfo{journal}{Phys. Rev. B}
  \textbf{\bibinfo{volume}{91}}, \bibinfo{pages}{235113}
  (\bibinfo{year}{2015}),
  \urlprefix\url{https://link.aps.org/doi/10.1103/PhysRevB.91.235113}.

\bibitem[{\citenamefont{Jager et~al.}(2017)\citenamefont{Jager, Ott, Kraus,
  Kaplan, Pouse, Marvel, Haglund, Neumark, and Leone}}]{Jager2017}
\bibinfo{author}{\bibfnamefont{M.~F.} \bibnamefont{Jager}},
  \bibinfo{author}{\bibfnamefont{C.}~\bibnamefont{Ott}},
  \bibinfo{author}{\bibfnamefont{P.~M.} \bibnamefont{Kraus}},
  \bibinfo{author}{\bibfnamefont{C.~J.} \bibnamefont{Kaplan}},
  \bibinfo{author}{\bibfnamefont{W.}~\bibnamefont{Pouse}},
  \bibinfo{author}{\bibfnamefont{R.~E.} \bibnamefont{Marvel}},
  \bibinfo{author}{\bibfnamefont{R.~F.} \bibnamefont{Haglund}},
  \bibinfo{author}{\bibfnamefont{D.~M.} \bibnamefont{Neumark}},
  \bibnamefont{and} \bibinfo{author}{\bibfnamefont{S.~R.} \bibnamefont{Leone}},
  \bibinfo{journal}{Proceedings of the National Academy of Sciences}
  \textbf{\bibinfo{volume}{114}}, \bibinfo{pages}{9558} (\bibinfo{year}{2017}),
  ISSN \bibinfo{issn}{0027-8424},
  \urlprefix\url{http://www.pnas.org/content/114/36/9558}.

\bibitem[{\citenamefont{Abreu et~al.}(2017)\citenamefont{Abreu, Gilbert~Corder,
  Yun, Wang, Ram\'{\i}rez, West, Zhang, Kittiwatanakul, Schuller, Lu
  et~al.}}]{Abreu2017}
\bibinfo{author}{\bibfnamefont{E.}~\bibnamefont{Abreu}},
  \bibinfo{author}{\bibfnamefont{S.~N.} \bibnamefont{Gilbert~Corder}},
  \bibinfo{author}{\bibfnamefont{S.~J.} \bibnamefont{Yun}},
  \bibinfo{author}{\bibfnamefont{S.}~\bibnamefont{Wang}},
  \bibinfo{author}{\bibfnamefont{J.~G.} \bibnamefont{Ram\'{\i}rez}},
  \bibinfo{author}{\bibfnamefont{K.}~\bibnamefont{West}},
  \bibinfo{author}{\bibfnamefont{J.}~\bibnamefont{Zhang}},
  \bibinfo{author}{\bibfnamefont{S.}~\bibnamefont{Kittiwatanakul}},
  \bibinfo{author}{\bibfnamefont{I.~K.} \bibnamefont{Schuller}},
  \bibinfo{author}{\bibfnamefont{J.}~\bibnamefont{Lu}}, \bibnamefont{et~al.},
  \bibinfo{journal}{Phys. Rev. B} \textbf{\bibinfo{volume}{96}},
  \bibinfo{pages}{094309} (\bibinfo{year}{2017}),
  \urlprefix\url{https://link.aps.org/doi/10.1103/PhysRevB.96.094309}.

\bibitem[{\citenamefont{Elsaesser and Woerner}(2010)}]{Elsaesser2010}
\bibinfo{author}{\bibfnamefont{T.}~\bibnamefont{Elsaesser}} \bibnamefont{and}
  \bibinfo{author}{\bibfnamefont{M.}~\bibnamefont{Woerner}},
  \bibinfo{journal}{Acta Crystallogr. Sect. A Found. Crystallogr.}
  \textbf{\bibinfo{volume}{66}}, \bibinfo{pages}{168} (\bibinfo{year}{2010}),
  ISSN \bibinfo{issn}{01087673}.

\bibitem[{SM()}]{SM}
\bibinfo{note}{See Supplemental Material at [URL] for more information}.

\bibitem[{\citenamefont{Lobastov et~al.}(2007)\citenamefont{Lobastov,
  Weissenrieder, Tang, and Zewail}}]{Lobastov2007}
\bibinfo{author}{\bibfnamefont{V.~A.} \bibnamefont{Lobastov}},
  \bibinfo{author}{\bibfnamefont{J.}~\bibnamefont{Weissenrieder}},
  \bibinfo{author}{\bibfnamefont{J.}~\bibnamefont{Tang}}, \bibnamefont{and}
  \bibinfo{author}{\bibfnamefont{A.~H.} \bibnamefont{Zewail}},
  \bibinfo{journal}{Nano Letters} \textbf{\bibinfo{volume}{7}},
  \bibinfo{pages}{2552} (\bibinfo{year}{2007}),
  \urlprefix\url{https://doi.org/10.1021/nl071341e}.

\bibitem[{\citenamefont{O{'}Callahan et~al.}(2015)\citenamefont{O{'}Callahan,
  Jones, Park, Cobden, Atkin, and Raschke}}]{ocallahan2015}
\bibinfo{author}{\bibfnamefont{B.~T.} \bibnamefont{O{'}Callahan}},
  \bibinfo{author}{\bibfnamefont{A.~C.} \bibnamefont{Jones}},
  \bibinfo{author}{\bibfnamefont{J.~H.} \bibnamefont{Park}},
  \bibinfo{author}{\bibfnamefont{D.~H.} \bibnamefont{Cobden}},
  \bibinfo{author}{\bibfnamefont{J.~M.} \bibnamefont{Atkin}}, \bibnamefont{and}
  \bibinfo{author}{\bibfnamefont{M.~B.} \bibnamefont{Raschke}},
  \bibinfo{journal}{Nature communications} \textbf{\bibinfo{volume}{6}},
  \bibinfo{pages}{6849} (\bibinfo{year}{2015}).

\bibitem[{\citenamefont{D\"onges et~al.}(2016)\citenamefont{D\"onges, Khatib,
  O’Callahan, Atkin, Park, Cobden, and Raschke}}]{Donges2016}
\bibinfo{author}{\bibfnamefont{S.~A.} \bibnamefont{D\"onges}},
  \bibinfo{author}{\bibfnamefont{O.}~\bibnamefont{Khatib}},
  \bibinfo{author}{\bibfnamefont{B.~T.} \bibnamefont{O’Callahan}},
  \bibinfo{author}{\bibfnamefont{J.~M.} \bibnamefont{Atkin}},
  \bibinfo{author}{\bibfnamefont{J.~H.} \bibnamefont{Park}},
  \bibinfo{author}{\bibfnamefont{D.}~\bibnamefont{Cobden}}, \bibnamefont{and}
  \bibinfo{author}{\bibfnamefont{M.~B.} \bibnamefont{Raschke}},
  \bibinfo{journal}{Nano Letters} \textbf{\bibinfo{volume}{16}},
  \bibinfo{pages}{3029} (\bibinfo{year}{2016}),
  \urlprefix\url{https://doi.org/10.1021/acs.nanolett.5b05313}.

\bibitem[{\citenamefont{Evans and Polanyi}(1935)}]{EyringPolyani}
\bibinfo{author}{\bibfnamefont{M.~G.} \bibnamefont{Evans}} \bibnamefont{and}
  \bibinfo{author}{\bibfnamefont{M.}~\bibnamefont{Polanyi}},
  \bibinfo{journal}{Trans. Faraday Soc.} \textbf{\bibinfo{volume}{31}},
  \bibinfo{pages}{875} (\bibinfo{year}{1935}),
  \urlprefix\url{http://dx.doi.org/10.1039/TF9353100875}.

\bibitem[{\citenamefont{He and Millis}(2016)}]{Millis2016}
\bibinfo{author}{\bibfnamefont{Z.}~\bibnamefont{He}} \bibnamefont{and}
  \bibinfo{author}{\bibfnamefont{A.~J.} \bibnamefont{Millis}},
  \bibinfo{journal}{Phys. Rev. B} \textbf{\bibinfo{volume}{93}},
  \bibinfo{pages}{115126} (\bibinfo{year}{2016}),
  \urlprefix\url{https://link.aps.org/doi/10.1103/PhysRevB.93.115126}.

\bibitem[{\citenamefont{Otto et~al.}(2017)\citenamefont{Otto, Ren\'{e}~de
  Cotret, Stern, and Siwick}}]{Otto2017}
\bibinfo{author}{\bibfnamefont{M.~R.} \bibnamefont{Otto}},
  \bibinfo{author}{\bibfnamefont{L.~P.} \bibnamefont{Ren\'{e}~de Cotret}},
  \bibinfo{author}{\bibfnamefont{M.~J.} \bibnamefont{Stern}}, \bibnamefont{and}
  \bibinfo{author}{\bibfnamefont{B.~J.} \bibnamefont{Siwick}},
  \bibinfo{journal}{Structural Dynamics} \textbf{\bibinfo{volume}{4}},
  \bibinfo{pages}{051101} (\bibinfo{year}{2017}),
  \urlprefix\url{https://doi.org/10.1063/1.4989960}.

\bibitem[{\citenamefont{Chatelain et~al.}(2012)\citenamefont{Chatelain,
  Morrison, Godbout, and Siwick}}]{Chatelain2012}
\bibinfo{author}{\bibfnamefont{R.~P.} \bibnamefont{Chatelain}},
  \bibinfo{author}{\bibfnamefont{V.~R.} \bibnamefont{Morrison}},
  \bibinfo{author}{\bibfnamefont{C.}~\bibnamefont{Godbout}}, \bibnamefont{and}
  \bibinfo{author}{\bibfnamefont{B.~J.} \bibnamefont{Siwick}},
  \bibinfo{journal}{Applied Physics Letters} \textbf{\bibinfo{volume}{101}},
  \bibinfo{pages}{081901} (\bibinfo{year}{2012}), ISSN
  \bibinfo{issn}{00036951},
  \urlprefix\url{http://link.aip.org/link/APPLAB/v101/i8/p081901/s1&Agg=doi}.

\bibitem[{\citenamefont{Liu et~al.}(2012)\citenamefont{Liu, Hwang, Tao,
  Strikwerda, Fan, Keiser, Sternbach, West, Kittiwatanakul, Lu
  et~al.}}]{liu2012}
\bibinfo{author}{\bibfnamefont{M.}~\bibnamefont{Liu}},
  \bibinfo{author}{\bibfnamefont{H.~Y.} \bibnamefont{Hwang}},
  \bibinfo{author}{\bibfnamefont{H.}~\bibnamefont{Tao}},
  \bibinfo{author}{\bibfnamefont{A.~C.} \bibnamefont{Strikwerda}},
  \bibinfo{author}{\bibfnamefont{K.}~\bibnamefont{Fan}},
  \bibinfo{author}{\bibfnamefont{G.~R.} \bibnamefont{Keiser}},
  \bibinfo{author}{\bibfnamefont{A.~J.} \bibnamefont{Sternbach}},
  \bibinfo{author}{\bibfnamefont{K.~G.} \bibnamefont{West}},
  \bibinfo{author}{\bibfnamefont{S.}~\bibnamefont{Kittiwatanakul}},
  \bibinfo{author}{\bibfnamefont{J.}~\bibnamefont{Lu}}, \bibnamefont{et~al.},
  \bibinfo{journal}{Nature} \textbf{\bibinfo{volume}{487}},
  \bibinfo{pages}{345} (\bibinfo{year}{2012}),
  \urlprefix\url{https://www.nature.com/articles/nature11231?page=3}.

\bibitem[{\citenamefont{Valverde-Chavez
  et~al.}(2015)\citenamefont{Valverde-Chavez, Ponseca, Stoumpos, Yartsev,
  Kanatzidis, Sundstrom, and Cooke}}]{ValverdeChavez2015}
\bibinfo{author}{\bibfnamefont{D.~A.} \bibnamefont{Valverde-Chavez}},
  \bibinfo{author}{\bibfnamefont{C.~S.} \bibnamefont{Ponseca}},
  \bibinfo{author}{\bibfnamefont{C.~C.} \bibnamefont{Stoumpos}},
  \bibinfo{author}{\bibfnamefont{A.}~\bibnamefont{Yartsev}},
  \bibinfo{author}{\bibfnamefont{M.~G.} \bibnamefont{Kanatzidis}},
  \bibinfo{author}{\bibfnamefont{V.}~\bibnamefont{Sundstrom}},
  \bibnamefont{and} \bibinfo{author}{\bibfnamefont{D.~G.} \bibnamefont{Cooke}},
  \bibinfo{journal}{Energy Environ. Sci.} \textbf{\bibinfo{volume}{8}},
  \bibinfo{pages}{3700} (\bibinfo{year}{2015}),
  \urlprefix\url{http://dx.doi.org/10.1039/C5EE02503F}.

\end{thebibliography}

\section{Methods}\label{sec:methods}

\subsection{Growth of vanadium dioxide films}
The VO$_2$ samples were deposited by pulsed laser deposition and grown to a thickness of 50~nm on top of a 40~nm SiN$_x$ substrate. The sample area is 250 $\mu$m by 250 $\mu$m and is formed by silicon aperture over the substrate. Details of the deposition process can be found in~\citet{Hendaoui2013}.

\subsection{RF-compressed ultrafast electron diffraction}

The ultrafast electron diffraction measurements are carried at $\sim300~$K in transmission mode with 90~keV RF-compressed electron pulses which have a bunch charge of $\sim0.1~$pC. The sample and the beamline are under high-vacuum ($\sim10^{-7}~$mBar). The compression cavity is driven by a $\sim2.998~$GHz ($40\times f_{\textup{rep-oscillator}}$) phase-locked harmonic tone of the oscillator pulse train and is detuning-compensated by active phase correction at 1 kHz~\cite{Otto2017}. The laser setup is based on a commercial Ti:Sapphire amplified system (Spectra-Physics Spitfire XP-Pro) operated at a repetition rate of 50--200~Hz (depending on excitation fluence) to allow for sufficient recovery of the VO$_2$ sample between laser pulses.  The duration of the optical pump pulse is 35 fs (fwhm) with a spot size of 350 $\mu$m (fwhm). The details and performance of the ultrafast electron diffraction instrument used in the work is described in detail elsewhere~\citep{Otto2017,Chatelain2012}.

\subsection{Time-resolved terahertz spectroscopy}
Time-resolved multi-terahertz spectroscopy experiments are based on two-color laser plasma generation of single cycle, broadband THz pulses and air-biased coherent detection providing a spectral range from 0.5--30 THz and temporal resolution of $\approx40$ fs. The THz spectrometer is driven by 35 fs duration, 795 nm pulses from an amplified Ti:sapphire femtosecond laser operating at 250 Hz repetition rate to allow for sample recovery between shots. The pump spot size was $\sim700~\mu$m which was at least 4 times the size of the THz pulse for the lowest frequencies analyzed here (2 THz). Experiments were performed in transmission at the focus of the THz pulse. The peak THz field strengths estimated to be $<50$ kV/cm, and under dry air purge gas conditions. At these field strengths no non-linear response of VO$_2$ is expected~\cite{liu2012}. Details of the experimental setup can be found in~\citet{ValverdeChavez2015}.

\section{Author Contributions}
B.J.S., D.G.C. and M.C. designed the experiments. M.R.O. and L.P.R. de C. performed the ultrafast electron diffraction experiments.  D.V.-C. and K.L.T. performed the THz spectroscopy measurements.  N.\'E. and M.C. fabricated the samples and performed the DC resistivity and XRD measurements.  M.R.O. and L.P.R. de C. analyzed the ultrafast electron diffraction data.  M.R.O., D.V.-C. and D.G.C. analyzed the THz data. M.R.O., L.P.R. de C., and B.J.S wrote the manuscript. M.R.O and L.P.R. de C. wrote the supplementary material. All authors contributed to revising and editing the manuscript.

\section{Competing interests}
The authors declare no competing financial interests.

\FloatBarrier

\end{document}